\documentclass[10pt,twocolumn,prd,aps,superscriptaddress,nofootinbib]{revtex4-1}
\usepackage[english]{babel}
\usepackage{amsmath}
\usepackage{amssymb}
\usepackage{mathrsfs}
\usepackage{graphicx}

\begin{document}

\title{Bounds on the polymer scale from gamma ray bursts}

\author{Yuri Bonder}
\email{bonder@nucleares.unam.mx}
\affiliation{Instituto de Ciencias Nucleares, Universidad Nacional Aut\'onoma de M\'exico\\
Apartado Postal 70-543, Ciudad de M\'exico 04510, M\'exico}

\author{Angel Garcia-Chung}
\email{angel.garcia@correo.nucleares.unam.mx}
\affiliation{Instituto de Ciencias Nucleares, Universidad Nacional Aut\'onoma de M\'exico\\
Apartado Postal 70-543, Ciudad de M\'exico 04510, M\'exico}
\affiliation{Departamento de F\'isica, Universidad Aut\'onoma Metropolitana - Iztapalapa\\
San Rafael Atlixco 186, Ciudad de M\'exico 09340, M\'exico}

\author{Saeed Rastgoo}
\email{saeed@xanum.uam.mx}
\affiliation{Departamento de F\'isica, Universidad Aut\'onoma Metropolitana - Iztapalapa\\
San Rafael Atlixco 186, Ciudad de M\'exico 09340, M\'exico}

\begin{abstract}
The polymer representations, which are partially motivated by loop quantum gravity, have been suggested as alternative schemes to quantize the matter fields. Here we apply a version of the polymer representations to the free electromagnetic field, in a reduced phase space setting, and derive the corresponding effective (i.e., semiclassical) Hamiltonian. We study the propagation of an electromagnetic pulse and we confront our theoretical results with gamma ray burst observations. This comparison reveals that the dimensionless polymer scale must be smaller than $4\times 10^{-35}$, casting doubts on the possibility that the matter fields are quantized with the polymer representation we employed.
\end{abstract}

\maketitle

Loop quantum gravity (LQG) \cite{LQProgram1, LQProgram2, LQProgram3}, which is a prominent quantum gravity candidate, has inspired alternative matter quantization methods, known as polymer representations \cite{Thiemann1,AshtekarLewansdoski,kaminski2006background,Viqar,Hugo}.
These alternative methods resemble LQG in that they are nonperturbative and unitarily inequivalent to the Schr\"odinger representation. Also, the formal way the states and the fundamental operators are expressed in the polymer representations, mimics the cylindrical functions and the holonomy-flux algebra of LQG, respectively. 
Moreover, the polymer representations have been considered, by themselves, as interesting alternatives to the Schr\"odinger quantization \cite{Emch,Strocchi,Halvorson,Strocchi2}.

Notably, most works on the polymer representations of matter fields use scalar fields or do not make contact with experimental data \cite{Varadarajan2,Varadarajan3,AbhayLew,seahra2012primordial,hassan2015polymer,Kajuri2015,garcia2016polymer}. In contrast, our goal is to study the empirical consequences of applying such a quantization scheme to the free electromagnetic field in the framework of Ref.~\onlinecite{Viqar}. To that end, we polymer quantize the Maxwell theory and then use well-known methods to extract the corresponding effective dynamics.

As it is well known, the electromagnetic field $A_{\nu}(x)$, $\nu$ being a spacetime index, has a $U(1)$ gauge symmetry and, to quantize it, we utilize a reduced phase space quantization (see, for example, Ref.~\onlinecite{Hanson}). 
Furthermore, we work in the Minkowski spacetime with a global Cartesian coordinate frame where $t$ represents the time index and $i,j$ are spatial indices. We fix the gauge by taking $A_t=0=\partial_i A^i$, which can be consistently imposed when there are no sources \cite[chapter 6.3]{Jackson}. In this case the action takes the form
\begin{equation}\label{action}
S=\frac{1}{2}\int dt\,d^{3}x\left[\partial_{t}A_{i}\partial_{t}A^{i}-\partial_{i}A_{j}\partial^{i}A^{j}\right].
\end{equation}
In this work we use a metric with signature $+2$ and adopt natural units, i.e., Lorentz-Heaviside units with the additional conditions $c=1=\hbar$. To get the Hamiltonian $H$, we use the spacetime foliation associated with constant $t$ hypersurfaces and denote the canonically conjugated momenta by $E^{i}$, resulting in
\begin{equation}
H= \frac{1}{2} \int d^{3}x \left[ E^{i}E_{i}+\partial_{i}A_{j}\partial^{i}A^{j}\right].\label{standard Hamiltonian}
\end{equation}
The fact that no constraints arise reflects that there is no remaining gauge freedom.

To properly implement the polymer quantization we turn to Fourier space. Notice, however, that \textit{a priori} we cannot assume Lorentz invariance, and thus, we do not use the standard four-dimensional Fourier transform. Instead we only perform such a transformation on the spatial coordinates. Furthermore, to have a countable number of modes, we consider the system to be in a finite box that induces an energy cutoff $\Lambda_{\rm c}$. Then, the fields can be written as 
\begin{subequations}\label{Fourier}
\begin{eqnarray}
A_{i}(\mathbf{x},t)&=&\sum_{\mathbf{k},r}\epsilon^{r}_{i}\left[\frac{1+i}{2}\mathscr{A}_r\left(\mathbf{k},t\right) + \frac{1-i}{2} \mathscr{A}_r \left(-\mathbf{k},t\right)\right]\nonumber\\&&\times e^{-i\mathbf{k}\cdot\mathbf{x}},\label{FTA}\\
E^{i}(\mathbf{x},t)&=&\sum_{\mathbf{k},r} \epsilon^{ir}\left[\frac{1+i}{2}\mathscr{E}_r\left(\mathbf{k},t\right) + \frac{1-i}{2} \mathscr{E}_r \left(-\mathbf{k},t\right)\right]\nonumber\\&&\times e^{-i\mathbf{k}\cdot\mathbf{x}},
\end{eqnarray}
\end{subequations}
where $\epsilon^{r}_{i}$ are the polarization vectors which satisfy $\epsilon^{r}_{i}k^i=0$, and the polarization index $r$ runs from $1$ to $2$. It can be checked that $\mathscr{A}_r$ and $\mathscr{E}_r$ are real, have mass dimensions $1$ and $2$, respectively, and are canonically conjugate, that is,
\begin{equation}
\left\{ \mathscr{A}_{r}\left(\mathbf{k},t\right),\mathscr{E}_{s}\left(\mathbf{k}^{\prime},t\right)\right\} =\Lambda_{\rm c}^3\delta_{rs} \delta(\mathbf{k}-\mathbf{k}^{\prime}),
\end{equation}
with all other Poisson brackets vanishing; $s$ is another polarization index. The factor $\Lambda_{\rm c}^3$ compensates for the fact that, in contrast to the Dirac delta, the Kronecker delta is dimensionless. In terms of these fields, the Hamiltonian \eqref{standard Hamiltonian} becomes
\begin{equation}
H= \frac{1}{2 \Lambda_{\rm c}^{3}}\sum_{\mathbf{k},r}\left[\mathscr{E}_{r}\left(\mathbf{k},t\right)^{2}+|\mathbf{k}|^{2}\mathscr{A}_{r}\left(\mathbf{k},t\right)^{2}\right],\label{HKModes}
\end{equation}
which has the form of a harmonic oscillator for each mode $\mathbf{k}$ and each polarization $r$.

We now implement the polymer representation on this classical theory in the spirit of Ref.~\onlinecite{Viqar}. We start by recalling that the Stone--von Neumann theorem \cite{StoneNeumann} states that, for any quantum system with finite degrees of freedom, any weakly continuous representation of the Weyl algebra is unitarily equivalent to the standard Schr\"odinger representation. There are situations, however, where the weak continuity assumption is not valid and the representation of the algebra is thus inequivalent to that of Schr\"odinger \cite{AshtekarWillis, CorichiZapata, Abhay}.

Now, to obtain the elements of the polymer quantization, it is convenient to first define the Weyl algebra for each mode $\mathbf{k}$. The generators of this algebra are denoted by $W(\mathscr{A}_1,\mathscr{A}_2,\mathscr{E}_1,\mathscr{E}_2)$ and their multiplication is given by
\begin{eqnarray}
&&W(\mathscr{A}_1,\mathscr{A}_2,\mathscr{E}_1,\mathscr{E}_2)W(\tilde{\mathscr{A}}_1,\tilde{\mathscr{A}}_2,\tilde{\mathscr{E}}_1,\tilde{\mathscr{E}}_2)\nonumber\\
&=&e^{\frac{i}{2}\Omega}W(\mathscr{A}_1+\tilde{\mathscr{A}}_1,\mathscr{A}_2+\tilde{\mathscr{A}}_2,\mathscr{E}_1+\tilde{\mathscr{E}}_1,\mathscr{E}_2+\tilde{\mathscr{E}}_2),
\label{multiplication}
\end{eqnarray} 
where $\Omega= \sum_{r=1,2} \left(\mathscr{E}_{r} \tilde{\mathscr{A}}_{r} - \mathscr{A}_{r}\tilde{\mathscr{E}}_{r}\right)/\Lambda_{\rm c}^{3}$ is the symplectic form evaluated at the corresponding phase-space point. This algebra can be used to define four groups by setting all but one of the arguments of $W$ to zero. The most relevant, for our purposes are 
\begin{eqnarray}\label{UV}
V_{1,\mathscr{E}_1}=W(0,0,\mathscr{E}_1,0),\quad
V_{2,\mathscr{E}_2}= W(0,0,0,\mathscr{E}_2).
\end{eqnarray}
Should the algebra representation be weakly continuous, there would be infinitesimal generators for all four groups defined above satisfying the canonical commutation relations. In our case, which is inspired by the holonomy-flux variables used in LQG, the weakly continuous condition of the Stone--von Neumann theorem is not satisfied, and thus, there are no infinitesimal generators for $V_{1,\mathscr{E}_1}$ and $V_{2,\mathscr{E}_2}$. Therefore, the fundamental operators are $\widehat{\mathscr{E}}_1$, $\widehat{\mathscr{E}}_2$, $\widehat{V}_{1,\mathscr{E}_1}$ and $\widehat{V}_{2,\mathscr{E}_2}$ which satisfy
\begin{equation}\label{CCR}
\left[\widehat{V}_{r,\mathscr{E}_r}, \widehat{\mathscr{E}}_s\right] = -\delta_{rs} \mathscr{E}_r \widehat{V}_{r,\mathscr{E}_r}.
\end{equation}

We now focus on one harmonic oscillator labeled by the fixed index $r$. The Hilbert space of such an oscillator is ${\cal H}^{(r)}_{poly} = L^2\left(\overline{\mathbb{R}}, d\mu_{Bohr}[{\mathscr{A}}_{r}]\right)$, where $\overline{\mathbb{R}}$ is the Bohr compactification of the real line and $d\mu_{Bohr}[{\mathscr{A}}_{r}]$ is the corresponding measure \cite{AshtekarWillis,Velhino}. Then, the wave functions can be expressed as almost periodic functions
\begin{equation}
\Psi({\mathscr{A}}_r) = \sum_{{\mathscr{E}}^{(n)}_r} \Psi_{{\mathscr{E}}^{(n)}_r} e^{-i \mathscr{E}^{(n)}_r \mathscr{A}_r/\Lambda_{\rm c}^{3}}, \label{PWF}
\end{equation}
with basis elements $e^{-i\mathscr{E}^{(n)}_r \mathscr{A}_r/\Lambda_{\rm c}^{3}}$. Such wave functions can be represented by a graph with a finite, but arbitrary, number of vertices $N$, with the $n$th vertex having a ``color'' ${\mathscr{E}}^{(n)}_r$, and $n=1,2,\dots,N$. Furthermore, the inner product with respect to the measure $d\mu_{Bohr}[{\mathscr{A}}_{r}]$ is 
\begin{equation}
\langle e^{-i{\mathscr{E}}^{(n)}_r {\mathscr{A}}_r/ \Lambda_{\rm c}^{3}}|e^{-i {\mathscr{E}}^{'(m)}_r {\mathscr{A}}_r / \Lambda_{\rm c}^{3}}\rangle=\delta_{ {\mathscr{E}}^{(n)}_r, {\mathscr{E}}^{'(m)}_r}.
\end{equation}
We emphasize that the right-hand side of the last equation is a Kronecker delta. Finally, the representation of the fundamental operators is
\begin{eqnarray}
\widehat{\mathscr{E}}_r \Psi({\mathscr{A}}_r) =& - i \Lambda_{\rm c}^{3} \frac{\delta}{\delta {\mathscr{A}}_r} \Psi({\mathscr{A}}_r), \\
\widehat{V}_{r,\mathscr{E}_r} e^{-i\mathscr{E}^{(n)}_r \mathscr{A}_r/\Lambda_{\rm c}^{3} }=& e^{-i(\mathscr{E}^{(n)}_r -\mathscr{E}_r)\mathscr{A}_r/\Lambda_{\rm c}^{3}},\label{Eq:VErepQ}
\end{eqnarray}
which correctly implements the commutators \eqref{CCR}.

The next step is to write the polymer quantum Hamiltonian. Our starting point is the classical Hamiltonian \eqref{HKModes} at a fixed time, so that, when promoted to an operator, it is in the Schr\"odinger representation. The fact that the operator $\widehat{\mathscr{A}}_r$ does not exist creates serious obstructions in representing the classical Hamiltonian, which depends on ${\mathscr{A}}_r^2$. This difficulty can be circumvented by replacing the operators $\widehat{{\mathscr{A}}}^2_r$ by a combination of Weyl generators. Specifically, we consider only regular graphs\footnote{For the polymer harmonic oscillator, the dynamics superselects equidistant graphs with polymer scale $\mu$ \cite{AshtekarWillis}. Moreover, when considering all possible shifts of a regular graph, the energy spectrum has a band structure \cite{Vergara,Barbero}. However, when $\mu$ is much smaller than the oscillator characteristic length, the bands' width is extremely narrow and it produces negligible physical effects.} that have equidistant values of ${\mathscr{E}}_r$, where the separation is given by a fixed, albeit arbitrary, positive parameter $\mu$, i.e., $\mathscr{E}^{(n)}_r=n\mu$. Note that we use the same $\mu$ for all polarizations and for every Fourier mode. This is a rather common assumption in this type of polymer quantization \cite{Viqar, Hugo} and $\mu$, known as the polymer scale, is thought to be of the order of the Planck scale (see, e.g., Ref.~\onlinecite{AshtekarWillis}). Concretely, we replace $\widehat{{\mathscr{A}}}_{r}^2$ in the Hamiltonian by
\begin{equation} \label{Apoly}
\widehat{{\mathscr{A}}}_{r}^2 \rightarrow \frac{\Lambda_{\rm c}^6}{\mu^2} \left[2 - \widehat{V}_{r,\mu} -{\widehat{V}_{r,-\mu}} \right],
\end{equation}
where, as can be seen from Eq.~\eqref{Eq:VErepQ}, $\widehat{V}_{r,\pm\mu}$, when applied to a basis element, produces a shift in $\mathscr{E}^{(n)}_r$ by $\pm\mu$. Note that, in the formal limit $\mu \to 0$, which only exists for regular representations, the right-hand side of Eq.~\eqref{Apoly} reduces to $\widehat{\mathscr{A}}_{r}^2 $. Under this replacement, the quantum polymer Hamiltonian associated with Eq.~\eqref{HKModes} becomes
\begin{equation}
\widehat{H} = \frac{1}{2 \Lambda_{\rm c}^{3}}\sum_{\mathbf{k},r} \left[ \widehat{\mathscr{E}}_{r}^2\left(\mathbf{k}\right) +\left(\frac{\Lambda_{\rm c}^3|\mathbf{k}|}{\mu} \right)^2 \left(2 - \widehat{V}_{r,\mu} - \widehat{V}_{r,-\mu} \right)\right].
\end{equation}

To derive the theoretical predictions that can be compared with available empirical data, we obtain the effective polymer Hamiltonian. This procedure is somehow technical, and it is thus described in Appendix \ref{appPoly} (see also Refs.~\cite{AshtekarWillis,Ashtekar:cam,Vergara,PRD92:104029,PRD95:065026,bojowald2006effective,singh2005semiclassical,ashtekar1999geometrical}). It turns out that such an effective Hamiltonian can be obtained by replacing
\begin{equation}\label{replacement}
\mathscr{A}_r\left(\mathbf{k}\right)^2 \rightarrow \left(\frac{2\Lambda_{\rm c}^{3}}{\mu} \right)^{2}\sin^{2}\left(\frac{\mu}{2\Lambda_{\rm c}^{3} } \mathscr{A}_r\left(\mathbf{k}\right)\right),
\end{equation}
in the classical action, which leads to the effective Hamiltonian
\begin{equation}
H_{\rm eff}= \frac{1}{2 \Lambda_{\rm c}^{3}}\sum_{\mathbf{k},r}\left[\mathscr{E}_{r}\left(\mathbf{k}\right)^{2}+ \left(\frac{2\Lambda_{\rm c}^{3} |\mathbf{k}| }{\mu} \right)^{2}\sin^{2}\left(\frac{\mu\mathscr{A}_r\left(\mathbf{k}\right)}{2\Lambda_{\rm c}^{3} } \right)\right].\label{Hsc}
\end{equation}
This Hamiltonian leads to the equations of motion
\begin{subequations}\label{effeom}
\begin{eqnarray}
\label{effeom1}\frac{d\mathscr{A}_r\left(\mathbf{k},t\right)}{dt}&=& \mathscr{E}_r\left(\mathbf{k},t\right),\\
\frac{d\mathscr{E}_r\left(\mathbf{k},t\right)}{dt}&=&- \frac{\Lambda_{\rm c}^{3}|\mathbf{k}|^2}{ \mu}\sin \left(\frac{\mu}{\Lambda_{\rm c}^{3}} \mathscr{A}_r\left(\mathbf{k},t\right)\right).
\end{eqnarray}
\end{subequations}
Equations~\eqref{effeom} are nonlinear, making it challenging to find wave solutions, and consequently the modified dispersion relations, as is typically done when looking for quantum gravity effects (cf. Ref.~\onlinecite{Viqar2009}). Still, we want to find empirical bounds on $\mu$, hence, we solve Eqs.~\eqref{effeom} perturbatively.

Note that standard electromagnetism is recovered from Eqs.~\eqref{effeom} when $\mu\to 0$, and since this theory properly describes all (classical) experiments, $\mu$ must be extremely small. We use this fact to solve Eqs.~\eqref{effeom} perturbatively where, to have a well defined perturbative expansion, we utilize the dimensionless polymer parameter $\tilde{\mu} = \mu/\Lambda_{\rm c}^{2}$. To obtain the perturbative equations it is convenient to first combine Eqs.~\eqref{effeom} into a single second-order equation for $\mathscr{A}_r\left(\mathbf{k},t\right)$, which, when expanded in $\tilde{\mu}$, takes the form
\begin{align}
0=&\partial^{2}_{t}{a}_r\left(\mathbf{k},t\right)+|\mathbf{k}|^2 a_r\left(\mathbf{k},t\right) +\tilde{\mu}^2\left[\partial^{2}_{t}\delta a_r\left(\mathbf{k},t\right)\right.\nonumber\\
& \left.+|\mathbf{k}|^2 \delta a_r\left(\mathbf{k},t\right)-\frac{|\mathbf{k}|^2}{6\Lambda_{\rm c}^{2}} a_r^3\left(\mathbf{k},t\right)\right]+O\left(\tilde{\mu}^4\right),\label{effeomexp}
 \end{align}
where $\mathscr{A}_r\left(\mathbf{k},t\right)=a_r\left(\mathbf{k},t\right)+\tilde{\mu}^2\delta a_r\left(\mathbf{k},t\right)+O(\tilde{\mu}^4)$. It can be verified that the solution to Eq.~\eqref{effeomexp} is
\begin{subequations}\label{solspert}
\begin{eqnarray}
a_r\left(\mathbf{k},t\right)&=& \mathscr{A}_r\left(\mathbf{k},0\right) \cos \left(|\mathbf{k}| t\right)+\frac{\mathscr{E}_r\left(\mathbf{k},0\right) }{|\mathbf{k}|} \sin \left(|\mathbf{k}|t\right),\nonumber\\ \\
\delta a_r\left(\mathbf{k},t\right)&=&\frac{|\mathbf{k}|}{6\Lambda_{\rm c}^{2}} \int_0^t ds\, a^3_r\left(\mathbf{k},s\right)\left[\sin \left(|\mathbf{k}| t\right) \cos \left(|\mathbf{k}| s\right)\right. \nonumber\\
 &&\left. -\cos \left(|\mathbf{k}| t\right)\sin \left(|\mathbf{k}| s\right)\right].
 \end{eqnarray}
 \end{subequations}

\begin{figure}[t]
\includegraphics[width=\columnwidth]{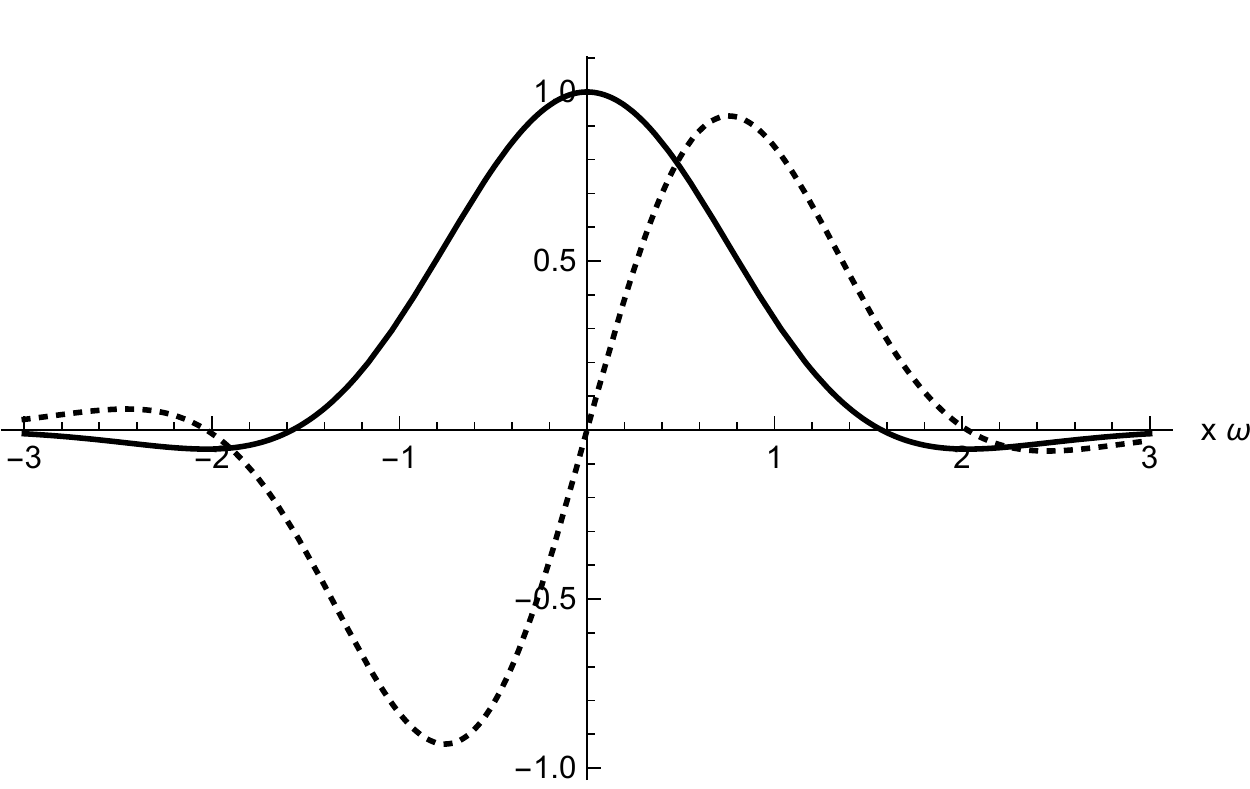}
\caption{\label{pulse}Initial pulse profile in units of $a$ for the particular case where $\sigma=\omega $. The solid and the dashed lines respectively represent $|\mathbf{A}(\mathbf{x},0)|$ and $|\mathbf{E}(\mathbf{x},0)|/\omega$.}
\end{figure}

We study the propagation of particular electromagnetic pulses according to Eqs.~\eqref{solspert}. Such pulses have been detected in the form of gamma ray bursts (GRBs), which are high-energy electromagnetic emissions from astrophysical sources that have played important roles in various quantum gravity phenomenology scenarios (see, for example, Refs.~\onlinecite{amelino,AlanMatt}). We model the GRB to be created in the form of a Gaussian pulse that propagates along the $x$ direction, oscillates transversely in the $y$ direction, and, at $t=0$, is centered at the origin and around the frequency $\omega$. Concretely, during an infinitesimal time interval around $t=0$, we take
\begin{equation}\label{initial pulse}
\mathbf{A}(\mathbf{x},t)=\mathbf{\hat{y}}a e^{-\sigma^{2}(x-t)^{2}/2}\cos[\omega(x-t)],
\end{equation}
where $a$ is the pulse amplitude and $\sigma$ is the Gaussian frequency width. The pulse's profile is plotted in Fig.~\ref{pulse} for the particular case where $\sigma=\omega$. We present the derivation of the solution $\mathbf{A}(\mathbf{x},t)$ for the initial data \eqref{initial pulse} in Appendix \ref{longcalc}.

\begin{figure}[t]
\includegraphics[width=\columnwidth]{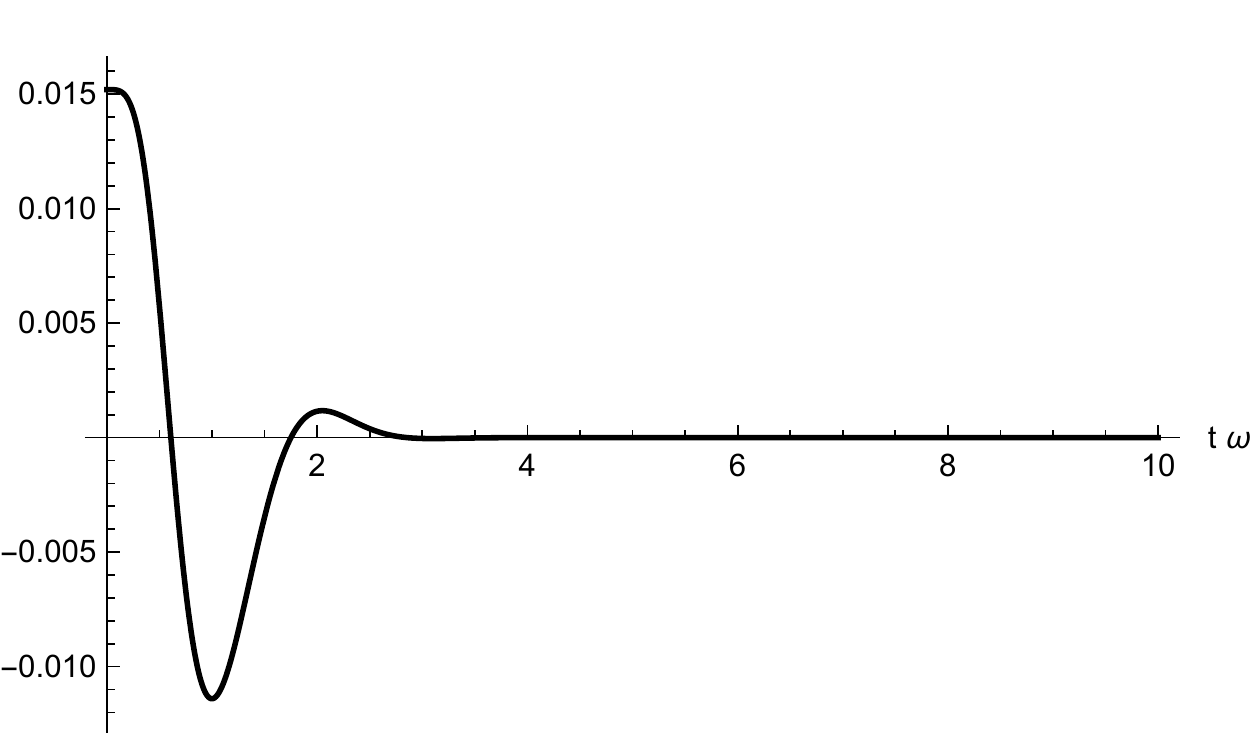}
\caption{\label{plotvel}Difference of the pulse speed and the large-time speed $v$ in units of $\tilde{\mu}^2 a^2 /\omega^2$ and as a function of $t \omega$, for the case where $\sigma=\omega$.}
\end{figure}

To compare with observations, we compute the pulse speed (in the frame we use throughout the paper). To define such a speed we follow the central pulse peak, since, as we mention above, there is no dispersion relation at our disposal from which we can read off a group velocity. We find this speed by using $x(t)=t+\tilde{\mu}^{2}\alpha(t)+ O(\tilde{\mu}^4)$ as an ansatz for the $x$ component of the central peak world line, and we determine $\alpha(t)$ by the conditions that the peak is an extremum of $|\mathbf{A}(\mathbf{x},t)|$, namely, that $\nabla\left|\mathbf{A}\left(\mathbf{x},t\right)\right|=\mathbf{0}$, and that, at $t=0$, this peak is centered at the origin (see Fig.~\ref{pulse}). We present the derivation of $\alpha(t)$ in Appendix \ref{longcalc}. Then, the pulse speed is simply $dx/dt$ and, as we also show in Appendix \ref{longcalc}, the $t$ dependence of $dx/dt$ drops as $e^{-2 \sigma^2 t^2/3}$, and thus, after a small time (with respect to $\sigma^{-1}$) the speed stabilizes to the large-time speed $v$ such that
\begin{equation}\label{deltav}
1-v=\frac{a^2 \tilde{\mu}^2 }{96 \sqrt{3}}\left(\frac{ \sigma^2+3 \omega^2+e^{-\frac{4 \omega^2}{3 \sigma^2}}\left(3 \sigma^2+\omega^2\right)}{ \sigma^2 \left(\sigma^2+\omega^2\right)}\right)+O\left(\tilde{\mu}^4\right).
\end{equation}
In Fig.~\ref{plotvel} we plot the difference of the pulse speed $dx/dt$ and the large-time speed $v$ as a function of $t$, for the particular case where $\sigma=\omega$. For $t\gtrsim 3 \omega$, such a difference becomes negligible as is also evident from this figure.
Thus, given that we are interested in comparing the theoretical predictions with astrophysical observations in which the time of flight is much larger than the time scales associated with $\sigma$, we neglect the time dependence of $dx/dt$ and take $v$ to describe the pulse speed. Still, the pulse speed depends on its frequency, frequency width, and amplitude.

We use the empirical data of a particular short GRB, known as GRB090510, which was detected by the GBM and LAT instruments onboard the Fermi Gamma-Ray Space Telescope \cite{GRB}. The GRB090510 event has $\omega\approx 30\ {\rm GeV}$ and the pulse energy ranges from hundreds of ${\rm keV}$ to tens of ${\rm GeV}$, setting the value of $\sigma$. Even though the pulse amplitude is not reported, from the total energy released by the GRB, we are able to infer that $a \approx 10^{28}\ {\rm GeV}$ (see Appendix \ref{appAmplitude}). Furthermore, it has been estimated that the traveling time difference for different frequencies satisfies $|\Delta t|< 859\ {\rm ms}$ \cite{bounds}. Since the traveling distance is $d\approx 10^{28}\ {\rm cm}$, we can conclude that the speed difference is restricted by $|1-v|< 3\times 10^{-18}$.

These experimental results can be compared with the the pulse speed prediction given by Eq.~\eqref{deltav}. The result is that, for the effective theory under consideration to properly describe the propagation of such a GRB,
\begin{equation}\label{bound}
\tilde{\mu} < 4 \times 10^{-35}.
\end{equation}
The stringency of this bound comes, mainly, from the enormous energy released by the GRB, and the large distance traveled by the light. There are studies in which more stringent limits on $\Delta t$ are set by combining several GRB observations \cite{betterbounds}, which would yield stronger bounds on $\tilde{\mu}$.

To get a sense of the stringency of the condition \eqref{bound}, we can set $\Lambda_{\rm c}^{-1}\sim D$, where $D\approx 10^{10}$ yr is the universe age \cite{universerasize}. That is, the size of the box in which we put the system to have a countable number of modes is of the order of the size of the observable universe. Under this assumption we get $\mu < 10^{-118}\ {\rm GeV}^2=10^{-156}\ l^2_{\rm P}/G^2$, where $G$ is Newton's constant and $l_{\rm P}$ is the Planck length. In other words, with this hypothesis, $\mu$ is restricted to be at least $156$ orders of magnitude below the expected scale.

To summarize, we have successfully applied a polymer quantization scheme to the free electromagnetic field in a fixed gauge. We then obtained the effective Hamiltonian, which leads to a nonlinear evolution and predicts that electromagnetic pulses propagate with subluminal speeds that depend on the pulses' frequency, frequency width, and amplitude. By comparing with the GRB data, we are able to conclude that, to reconcile the theory with observations, the polymer scale $\mu$, when divided by the cutoff scale squared, has to be smaller than $4\times 10^{-35}$. We would like to stress that although other studies \cite{Hugo,husain2016low,Kajuri2016} have found obstructions on alternative matter polymer representations, our analysis is the first to use physical fields, to actually connect the predictions with existing observations, and to put a bound on the polymer scale. Importantly, the strong bound we set suggests that the polymer representation that we employed may not be directly related with the presumed quantum gravity scale, and that the method under consideration may not be the way the matter fields in nature are quantized. 

Finally, an interesting extension of our work which could shed light into the quantum nature of spacetime itself is to study the gravitational waves in the effective polymer description, particularly since there are experimental constraints on the speed of such waves \cite{LIGO,LIGO2}. In addition, it would be enlightening to study the behavior of the electromagnetic constraints when the field is quantized polymerically, in which case, one cannot fix the gauge at the classical level.

\begin{acknowledgments}
We thank V. Husain, H. Morales-T\'ecotl, D. Sudarsky, and D. Vergara for useful discussions. A. G. C. thanks the University of New Brunswick's Gravity Group for their feedback and hospitality. We acknowledge the support from UNAM-DGAPA-PAPIIT Grants No.  IA101116 and No. IA101818 (Y. B.) and No. IN103716 (A. G. C.), CONACyT Grants No. 237351 (S. R., A. G. C.) and No. 237503 (A. G. C.), UNAM-DGAPA postdoctoral fellowship (A. G. C.), and Red FAE CONACyT.\end{acknowledgments}

\begin{widetext}
\appendix
\section{Effective dynamics}\label{appPoly}

In this appendix we derive the effective dynamics for the theory under consideration. We first study the polymer amplitude and we then take the continuum limit. At this point it is possible to extract the semiclassical action, where the replacement \eqref{replacement} can be justified.

The polymer amplitude satisfies
\begin{equation}
\langle \mathscr{A}_{r, f},t_f |\mathscr{A}_{r, i},t_i \rangle=
 \langle \mathscr{A}_{r, f} | e^{-i (t_f - t_i) \widehat{H}} |\mathscr{A}_{r, i} \rangle
= \left[ \prod^N_{n=1} \int^{+\pi \mathscr{A}^c}_{-\pi \mathscr{A}^c} \frac{d \mathscr{A}_{r,n} }{2 \pi \mathscr{A}^c} \right] \prod^{N+1}_{k=1} \langle \mathscr{A}_{r,k} , t_k | \mathscr{A}_{r, k-1} , t_{k-1} \rangle, \label{FeyAmp}
\end{equation}
where $\mathscr{A}^c=\Lambda^3_c/\mu$ and $\epsilon = t_k - t_{k-1}$ are infinitesimal. This calculation cannot be done using the conventional techniques since the $\mathscr{E}_r$ take values in discrete sets, which implies that the $\mathscr{A}_r$ are compact and satisfy $ (1/2 \pi \mathscr{A}^c) \int^{+\pi \mathscr{A}^c}_{-\pi \mathscr{A}^c} d \mathscr{A}_{r} | \mathscr{A}_{r} \rangle \langle \mathscr{A}_{r} | = 1$.

As it is done in Ref. \cite[chapter 6.1]{Kleinert}, we first compute the infinitesimal amplitude for a vanishing Hamiltonian
\begin{eqnarray}
\langle \mathscr{A}_{r,k} , t_k | \mathscr{A}_{r, k-1} , t_{k-1} \rangle^{(0)}& =& \langle \mathscr{A}_{r,k}| \mathscr{A}_{r, k-1}\rangle \nonumber\\
&=& 2\pi \mathscr{A}^c \sum_{n \in \mathbb{Z} } \delta\left( \mathscr{A}_{r, k} - \mathscr{A}_{r, k-1} - 2 \pi n \mathscr{A}^c \right) \nonumber \\
&=& \frac{1}{2} \sum_{n \in \mathbb{Z}} \int^{+\infty}_{-\infty} d\varphi_k e^{ i \varphi_k \left( \mathscr{A}_{r,k} - \mathscr{A}_{r,k-1} - 2\pi n \mathscr{A}^c \right) /2 \mathscr{A}^c},\label{zeroHamAmp}
\end{eqnarray}
where $\varphi_k$ are auxiliary variables. Then, to calculate $\langle \mathscr{A}_{r,k} , t_k | \mathscr{A}_{r, k-1} , t_{k-1} \rangle$ we follow the derivation given in chapter 2.1 of Ref. \cite[pages 89-94]{Kleinert}, which calls for the amplitude \eqref{zeroHamAmp}. The result is
\begin{equation}
 \langle \mathscr{A}_{r,k} , t_k | \mathscr{A}_{r, k-1} , t_{k-1} \rangle =\sum_{n_k \in \mathbb{Z}} \int^{+\infty}_{-\infty} \frac{d\varphi_k}{2} e^{ \frac{i \varphi_k}{2 \mathscr{A}^c} \left( \mathscr{A}_{r,k} - \mathscr{A}_{r,k-1} - 2\pi n_k \mathscr{A}^c \right) -i \epsilon H^{(k)}},\nonumber\\ \label{ZeroAmp}
\end{equation}
where $H^{(k)}$ is the Hamiltonian \eqref{HKModes} evaluated at $\mathscr{A}_{r}=2\mathscr{A}^c \sin\left( \mathscr{A}_{r,k}/2\mathscr{A}^c \right) $ and $\mathscr{E}_{r}=\mu \varphi_k/2$. 

Next we substitute Eq. \eqref{ZeroAmp} in the amplitude \eqref{FeyAmp} and redefine the integration variables $\mathscr{A}_{r,n_k} \rightarrow \mathscr{A}_{r,n_k} - 2 \pi n_k \mathscr{A}^c$, which leaves only one sum (more details on this last step can be found in Ref. \onlinecite[Appendix B]{Vergara}). Then, the amplitude \eqref{FeyAmp} takes the form
\begin{equation}
\langle \mathscr{A}_{r, f},t_f |\mathscr{A}_{r, i} ,t_i \rangle = \sum_l \left[ \prod^N_{n=1} \int^{+\infty}_{-\infty} \frac{d \mathscr{A}_{r,n} }{2 \pi \mathscr{A}^c} \right] \left[ \prod^{N+1}_{k=1} \int^{+\infty}_{-\infty} \frac{d\varphi_k}{2}\right] e^{ \sum^{N+1}_{k=1} \left[ i \varphi_k \left( \mathscr{A}_{r,k} - \mathscr{A}_{r,k-1} - 2\pi l \delta_{k,N+1} \mathscr{A}^c \right)/2 \mathscr{A}^c -i \epsilon H^{(k)} \right] }. \label{amplarga}
\end{equation}
After integrating the auxiliary variables, the right-hand side of Eq.~\eqref{amplarga} becomes
\begin{equation}
 \sum_l \left[ \prod^N_{n=1} \int^{+\infty}_{-\infty} \frac{d \mathscr{A}_{r,n} }{2 \pi \mathscr{A}^c} \right] \prod^{N+1}_{k=1} \sqrt{\frac{2 \pi}{i \epsilon \mu^2} } \exp \left[ - \frac{ \left( \mathscr{A}_{r,k} - \mathscr{A}_{r,k-1} - 2\pi l \delta_{k,N+1} \mathscr{A}^c \right)^2}{2i \Lambda^3_c (t_k - t_{k-1})} - \frac{2i (t_k - t_{k-1})|\mathbf{k}|^2 (\mathscr{A}^c)^2}{\Lambda^3_c} \sin^2\left( \frac{ \mathscr{A}_{r,k}}{2 \mathscr{A}^c} \right) \right].\label{RHS}
\end{equation}
The last step is to take the continuum limit $N\to \infty$ in Eq. \eqref{RHS}, which implies
\begin{equation}
\langle \mathscr{A}_{r, f},t_f |\mathscr{A}_{r, i},t_i \rangle = \sum_l \int^{\mathscr{A}_{r, f} + 2 \pi l \mathscr{A}^c}_{\mathscr{A}_{r,i} } \frac{{\cal D} \mathscr{A}_{r}}{2 \pi \mathscr{A}^c} e^{ iS_{\rm eff}/ \Lambda^3_c }, 
\end{equation}
where ${\cal D} \mathscr{A}_{r}$ is the formal notation for the measure and the effective action is
\begin{equation}
S_{\rm eff} = \int^{t_f}_{t_i} dt \left[ \frac{1}{2} \dot{\mathscr{A}}^2_{r} - \frac{|\mathbf{k}|^2}{2} \left( \frac{2 \Lambda^3_c}{\mu} \right)^2 \sin^2\left( \frac{\mu \mathscr{A}_{r}}{2 \Lambda^3_c} \right)\right].
\end{equation}
Observe that this effective action can be obtained from Eq.~\eqref{action} after making the replacement \eqref{replacement}. This result justifies such a replacement as a method to get the effective limit from the polymer quantum theory. We want to emphasize that our derivation was possible because the field modes are described by quantum harmonic oscillators and, in this case, there are no ambiguities; in other theories one needs to be careful when applying similar replacements.

\section{Detailed phenomenological analysis}\label{longcalc}

Here we present the computational details of some of the results of the phenomenological part of the paper. We first focus on the expression for $\mathscr{A}_r\left(\mathbf{k},t\right)$ that is a solution with the initial data \eqref{initial pulse}. To put these data in the form required by the solution \eqref{solspert}, we use the inverse of Eq.~\eqref{FTA} and Eq. \eqref{effeom1}; the result is
\begin{subequations}\label{initialdata}
\begin{eqnarray}
\mathscr{A}_1\left(\mathbf{k},0\right)&=&\frac{a \Lambda_{\rm c} }{2 \sigma }\left( e^{-\frac{(|\mathbf{k}| -\omega )^2}{2 \sigma^2}}+
 e^{-\frac{(-|\mathbf{k}| -\omega )^2}{2 \sigma^2}}\right),\\
 \mathscr{A}_2\left(\mathbf{k},0\right)&=&0,\\
\mathscr{E}_r\left(\mathbf{k},0\right)&=&|\mathbf{k}| \mathscr{A}_r\left(\mathbf{k},0\right).
 \end{eqnarray}
 \end{subequations}
We then insert these initial conditions into Eqs. \eqref{solspert}, which, after some simplifications, lead to $\mathscr{A}_2\left(\mathbf{k},t\right)=0$ and
\begin{eqnarray}
\mathscr{A}_1\left(\mathbf{k},t\right)&=&\frac{a \Lambda_c}{2 \sigma } \left(e^{-\frac{(|\mathbf{k}|+\omega )^2}{2 \sigma^2}}+e^{-\frac{(-|\mathbf{k}|+\omega )^2}{2 \sigma^2}}\right) \left[\sin (|\mathbf{k}| t)+\cos (|\mathbf{k}| t)\right]
+\tilde{\mu}^2\frac{a^3 \Lambda_c}{768 \sigma^3} \left(e^{-\frac{(|\mathbf{k}|+\omega )^2}{2 \sigma^2}}+e^{-\frac{(-|\mathbf{k}|+\omega )^2}{2 \sigma^2}}\right)^3\nonumber\\
&&\times \left\{\cos (3 |\mathbf{k}| t)-(12 |\mathbf{k}| t+1) \cos (|\mathbf{k}| t)+\left[12 |\mathbf{k}| t-2 \cos (2 |\mathbf{k}| t)+14\right]\sin (|\mathbf{k}| t) \right\}+O\left(\mu^4\right).
\end{eqnarray}
It can easily be verified that, at $t=0$, these expressions reduce to Eqs. \eqref{initialdata}. Next, we use Eq.~\eqref{FTA} to derive
\begin{eqnarray}
\mathbf{A}\left(\mathbf{x},t\right)&=&\mathbf{\hat{y}} a e^{-\frac{1}{2} \sigma^2 (t-x)^2} \cos [\omega (t-x)]
+\tilde{\mu}^2\frac{\mathbf{\hat{y}} a^3 }{384 \sqrt{3} \sigma^2} e^{(-9 t^2 \sigma^4 - x^2 \sigma^4 - 6 t x \sigma^4 - 
 8 \omega^2)/(6\sigma^2)}\nonumber\\
 && \times \left\{3 \left(4 \sigma^2 t^2-4 \sigma^2 t x+7\right)e^{4\sigma^2 t (t+x)/3} \cos \left[\frac{\omega}{3} (t-x)\right]
 +12 t \omega e^{4 \left(\sigma^4 t^2+\sigma^4 t x+\omega^2\right)/(3 \sigma^2)} \sin [\omega (t-x)] \right. \nonumber\\
 && +\left(4 \sigma^2 t^2-4 \sigma^2 t x+7\right) e^{4 \left(\sigma^4 t^2+\sigma^4 t x+\omega^2\right)/(3 \sigma^2)} \cos [\omega (t-x)]
 -8 e^{\left(4 \sigma^4 t^2+2\sigma^4 t x+4 \omega^2\right)/(3 \sigma^2)} \cos [\omega(t+x)]
 \nonumber\\
 &&+e^{4 \omega^2/(3 \sigma^2)} \cos [\omega (3 t+x)]+12 t \omega e^{4 \sigma^2 t (t+x)/3} \sin \left[\frac{\omega}{3} (t-x)\right]-24 e^{2 \sigma^2 t (2 t+x)/3} \cos \left[\frac{\omega}{3} (t+x)\right] \nonumber\\
 &&\left.+3 \cos \left(t \omega +\frac{x \omega }{3}\right)\right\} +O\left(\mu^4\right).\label{Avecsol}
 \end{eqnarray}
Because of the complicated form of the above expression, it is hard to do a full consistency check, however, we can verify that the above equation reduces to the corresponding initial data at $t=0$.

We now want to find the propagation speed of the central peak, which is the physical quantity we use to compare with the experimental observations. We employ the ansatz $x(t)=t+\tilde{\mu}^{2}\alpha(t)+O(\tilde{\mu}^{4})$ for the central peak world line. The value of $\alpha(t)$ can be found using that the central peak is an extremum of $\left|\mathbf{A}\left(\mathbf{x},t\right)\right|$, i.e., it satisfies $\nabla\left|\mathbf{A}\left(\mathbf{x},t\right)\right|=\mathbf{0}$. When we take the gradient of the norm of Eq.~\eqref{Avecsol} and evaluate it at $x(t)=t+\tilde{\mu}^{2}\alpha(t)$, we get that, at order $O(\tilde{\mu}^{0})$,
\begin{equation}
\nabla\left|\mathbf{A}\left(\mathbf{x},t\right)\right|_{\tilde{\mu}=0}=\mathbf{\hat{x}}ae^{-\sigma^{2}(t-x)^{2}/2}\left\{ \sigma^{2}(t-x)\cos[\omega(t-x)]+\omega\sin[\omega(t-x)]\right\}.\label{convsol}
\end{equation}
This last equation clearly vanishes for $\left.x(t)\right|_{\tilde{\mu}=0}=t$, recovering the well-known result that, according to conventional electrodynamics, pulses propagate at the speed of light. The $O(\tilde{\mu}^{2})$ contribution has two parts: one from evaluating $\nabla\left|\mathbf{A}\left(\mathbf{x},t\right)\right|_{\tilde{\mu}=0}$ at $\tilde{\mu}^{2}\alpha(t)$, and a second from the $O(\tilde{\mu}^{2})$ part of $\nabla\left|\mathbf{A}\left(\mathbf{x},t\right)\right|$, which is evaluated at $\left.x(t)\right|_{\tilde{\mu}=0}=t$. From setting the resulting expression to zero we obtain
\begin{eqnarray}
\alpha(t)&=&\frac{-a^2 }{1152 \sqrt{3} \sigma^2 \left(\sigma^2+\omega^2\right)}
   \left\{12t (\sigma^2 +3\omega^2)+12 t (\omega^2 +3 \sigma^2) e^{-4 \omega^2/(3 \sigma^2)} -8 e^{-2 \sigma^4 t^2/(3 \sigma^2)}\left[3 \omega\sin (2 t \omega )+2 \sigma^2 \cos (2 t \omega )\right]\right.\nonumber\\
   &&-24  e^{(-2 \sigma^4 t^2-4 \omega^2)/(3 \sigma^2)}\left[ \omega \sin \left(\frac{2 t \omega
   }{3}\right)+2 \sigma^2t \cos \left(\frac{2 t \omega  }{3}\right)\right]+  e^{-8 \sigma^4 t^2/(3 \sigma^2)}\left[3 \omega\sin (4 t \omega )+4 \sigma^2 t \cos (4 t \omega )\right]\nonumber\\
  &&\left. +3 e^{(-8 \sigma^4 t^2-4 \omega^2)/(3 \sigma^2)}\left[4\sigma^2 t \cos \left(\frac{4 t   \omega }{3}\right)+3 \omega  \sin \left(\frac{4 t \omega   }{3}\right)\right]\right\}.\label{alphalong}
\end{eqnarray}
It can be directly verified that $\alpha(0)=0$, and therefore, $x(0)=O(\tilde{\mu}^4)$, which ensures that we follow the central peak of the pulse and not another extremum of $\left|\mathbf{A}\left(\mathbf{x},t\right)\right|$. Importantly, the fact that there exists a solution of $\alpha(t)$ for all $t$, shows that within our perturbative approach, the central peak can be traced for all times. Whether such a peak can be traced using the unperturbed dynamics given in Eqs.~\eqref{effeom} is an open question that is left to a future analytical or numerical study. 

Finally, the speed of the pulse's central peak is
\begin{eqnarray}
\frac{dx}{dt}&=&1+ \tilde{\mu}^2 \frac{d \alpha(t)}{dt}+O(\tilde{\mu}^4)\nonumber\\
&=&1-\frac{\tilde{\mu}^2a^2}{864 \sqrt{3} \sigma^2 \left(\sigma^2+\omega ^2\right)}
\left\{9 (\sigma^2+3 \omega^2)+9 (\omega^2 +3 \sigma ^2) e^{-4\omega ^2/(3\sigma ^2)}\right.\nonumber\\
&&+4 e^{-2 \sigma^2 t^2/3} [12t \omega\sigma^2 \sin (2 t \omega)+(4\sigma^4 t^2-3\sigma^2-9 \omega^2)  \cos   (2 t \omega )] \nonumber\\
&&+12   e^{(-2 \sigma ^4 t^2-4\omega ^2)/(3\sigma ^2)} \left[4 \sigma^2 t \omega\sin \left(\frac{2 t \omega }{3}\right) +\left(4 \sigma ^4 t^2-3 \sigma ^2-\omega ^2\right) \cos\left(\frac{2 t \omega }{3}\right)\right]\nonumber\\
&&+e^{-8 \sigma ^4 t^2/(3\sigma ^2)}\left[-24 \sigma^2 t \omega   \sin (4 t \omega )+(-16 \sigma^4 t^2+3 \sigma^2+9 \omega^2 ) \cos (4 t \omega )\right]\nonumber\\
&&\left.+ e^{(-8 \sigma ^4 t^2-4\omega ^2)/(3\sigma ^2)}  \left[ -24 \sigma^2 t \omega\sin \left(\frac{4 t \omega }{3}\right) +\left(9 \sigma ^2-48 \sigma ^4 t^2+3 \omega^2\right)  \cos \left(\frac{4 t \omega }{3}\right)\right]\right\}+O\left(\tilde{\mu}^4\right).
\end{eqnarray}
Clearly, in the limit $t \sigma\gg 1$, the last four lines are exponentially suppressed, and we get the large-time speed $v$ given in Eq. \eqref{deltav}.

\section{GRB amplitude}\label{appAmplitude}

The goal of this appendix is to infer the value of the pulse amplitude $a$ from the reported data: the pulse's frequency, frequency width, and total released energy, which has been estimated at $U\approx 6\times 10^{55}\ {\rm GeV}$ \cite{energy}. This part of the analysis can be done using standard electromagnetism, since, in Eq.~\eqref{deltav}, $a$ is suppressed by $\tilde{\mu}^2$, and thus, any additional $\tilde{\mu}$ correction lies at the order we neglect. Moreover, we assume that the GRB is well described by a three-dimensional spherical Gaussian pulse (since we are only looking for an order-of-magnitude estimation, we ignore that spherical symmetric systems do not radiate). Around the emission time $t=0$, such a pulse can be described by
\begin{equation}
\mathbf{A}(\mathbf{x},t)=a\mathbf{\hat{\phi}} e^{-\sigma^{2}(r-t)^{2}/2}\cos[\omega(r-t)],
\end{equation}
where we use conventional spherical coordinates $r$, $\theta$, and $\phi$. This field is divergence free and, importantly, $a$, $\omega$, and $\sigma$ play the same roles as in Eq.~\eqref{initial pulse}.

The total energy of an electromagnetic configuration is given by the Hamiltonian \eqref{standard Hamiltonian} \cite[chapter 6.7]{Jackson}. This total energy for the pulse under consideration [taking into the account that Eq.~\eqref{standard Hamiltonian} is written in Cartesian coordinates], at $t=0$, is
\begin{equation}
U =\frac{\pi^{3/2} a^2}{4 \sigma^3}\left[2 \omega^2 \left(1-e^{-\frac{\omega^2}{\sigma^2}}\right) + (3+\ln 4)\sigma^2 \left(1+e^{-\frac{\omega^2}{\sigma^2}}\right)\right].
\end{equation}
Using the particular values for the GRB under consideration (see the text for the values of $\omega$ and $\sigma$; we neglect the frequency shift due to the relative speed of the source and the detector) we find $a \approx 10^{28}\ {\rm GeV}$, which is the quantity we require.
\end{widetext}

\bibliography{main}

\end{document}